\documentclass[aps,pra,superscriptaddress]{revtex4}
\usepackage{amsmath,amssymb,epsfig}
\usepackage{graphicx}

\def\be{\begin{equation}}
\def\ee{\end{equation}}
\def\bea{\begin{eqnarray}}
\def\eea{\end{eqnarray}}

\newcommand{\avg}[1]{\mbox{$\langle#1\rangle$}}

\def\kappaex{\kappa_{\footnotesize\textrm{ex}}}

\def\kappae{\kappaex}

\newcommand{\opdagger}[2]{\mbox{$\hat{#1}_{#2}^{\dagger}$}}
\newcommand{\op}[2]{\mbox{$\hat{#1}_{#2}$}}
\def\npump{\avg{n_\text{c}}}
\def\DeltaOP{\Delta_\text{OC}}
\def\omegap{\omega_\text{c}}
\def\tpd{\Delta_{\text{SC}}}
\newcommand{\m}[1]{\mbox{$\mathbf{#1}$}}

\begin{document}
\title{Electromagnetically Induced Transparency and Slow Light with Optomechanics}
\date{\today}

\author{A. H. Safavi-Naeini}\thanks{These authors
contributed equally to this work.} \affiliation{Thomas J. Watson, Sr., Laboratory of Applied Physics,
California Institute of Technology, Pasadena, CA 91125}
\author{T. P. Mayer Alegre}\thanks{These authors contributed equally to this work.}
\affiliation{Thomas J. Watson, Sr., Laboratory of Applied Physics,
California Institute of Technology, Pasadena, CA 91125}
\author{J. Chan}
\affiliation{Thomas J. Watson, Sr., Laboratory of Applied Physics,
California Institute of Technology, Pasadena, CA 91125}
\author{M. Eichenfield}
\affiliation{Thomas J. Watson, Sr., Laboratory of Applied Physics,
California Institute of Technology, Pasadena, CA 91125}
\author{M. Winger}
\affiliation{Thomas J. Watson, Sr., Laboratory of Applied Physics,
California Institute of Technology, Pasadena, CA 91125}
\author{Q. Lin}
\affiliation{Thomas J. Watson, Sr., Laboratory of Applied Physics,
California Institute of Technology, Pasadena, CA 91125}
\author{J. T. Hill}
\affiliation{Thomas J. Watson, Sr., Laboratory of Applied Physics,
California Institute of Technology, Pasadena, CA 91125}
\author{D. E. Chang}\affiliation{Institute for
Quantum Information and Center for the Physics of Information,
California Institute of Technology, Pasadena, CA 91125}

\author{O. Painter}
\affiliation{Thomas J. Watson, Sr., Laboratory of Applied Physics, California Institute of Technology, Pasadena, CA 91125}

\begin{abstract}
Controlling the interaction between localized optical and mechanical excitations has recently become possible following advances in micro- and nano-fabrication techniques~\cite{Kippenberg2008,Favero2009}.  To date, most experimental studies of optomechanics have focused on measurement and control of the mechanical subsystem through its interaction with optics, and have led to the experimental demonstration of dynamical back-action cooling and optical rigidity of the mechanical system~\cite{Braginsky1977,Kippenberg2008}.  Conversely, the optical response of these systems is also modified in the presence of mechanical interactions, leading to strong nonlinear effects such as Electromagnetically Induced Transparency (EIT)~\cite{Weis2010} and parametric normal-mode splitting~\cite{groblacher09a}. In atomic systems, seminal experiments~\cite{Hau1999} and proposals to slow and stop the propagation of light~\cite{fleischhauer05}, and their applicability to modern optical networks~\cite{Boyd2009}, and future quantum networks~\cite{Kimble2008}, have thrust EIT  to the forefront of experimental study during the last two decades. In a similar fashion, here we use the optomechanical nonlinearity to control the velocity of light via engineered photon-phonon interactions. Our results demonstrate EIT and tunable optical delays in a nanoscale optomechanical crystal device, fabricated by simply etching holes into a thin film of silicon (Si). At low temperature ($8.7$~K), we show an optically-tunable delay of $50~\text{ns}$ with near-unity optical transparency,  and superluminal light with a $1.4~\mu \text{s}$ signal advance. These results, while indicating significant progress towards an integrated quantum optomechanical memory~\cite{Chang2010}, are also relevant to classical signal processing applications.  Measurements at room temperature and in the analogous regime of Electromagnetically Induced Absorption (EIA) show the utility of these chip-scale optomechanical systems for optical buffering, amplification, and filtering of microwave-over-optical signals.
\end{abstract}
\maketitle

It is by now well known that the optical properties of matter can be dramatically modified by using a secondary light beam, approximately resonant with an internal process of the material system. As an example, an opaque object can be made transparent in the presence of a control beam, in what is referred to as ``Electromagnetically Induced Transparency'' (EIT). A remarkable feature of EIT is the drastic reduction in the group velocity of light passing through the material, achieved inside a practically lossless transparency window. This aspect of the effect has been utilized to conjure schemes whereby light may be slowed and stopped, making it an important building block in quantum information and communication proposals, as well as of great practical interest in classical optics and photonics. A simple upper-bound for the storage time in EIT-based proposals is the lifetime related to the internal processes of the material. These lifetimes can be extremely long in atomic gases, with storage times on the order of one second having been demonstrated~\cite{Zhang2009} in Bose-Einstein Condensates. Part of the vision for future scalable quantum networks has involved extending the remarkable results achieved in atomic experiments to a more readily deployable domain.

\begin{figure*}[ht!]
\begin{center}
\includegraphics[width=17.5cm]{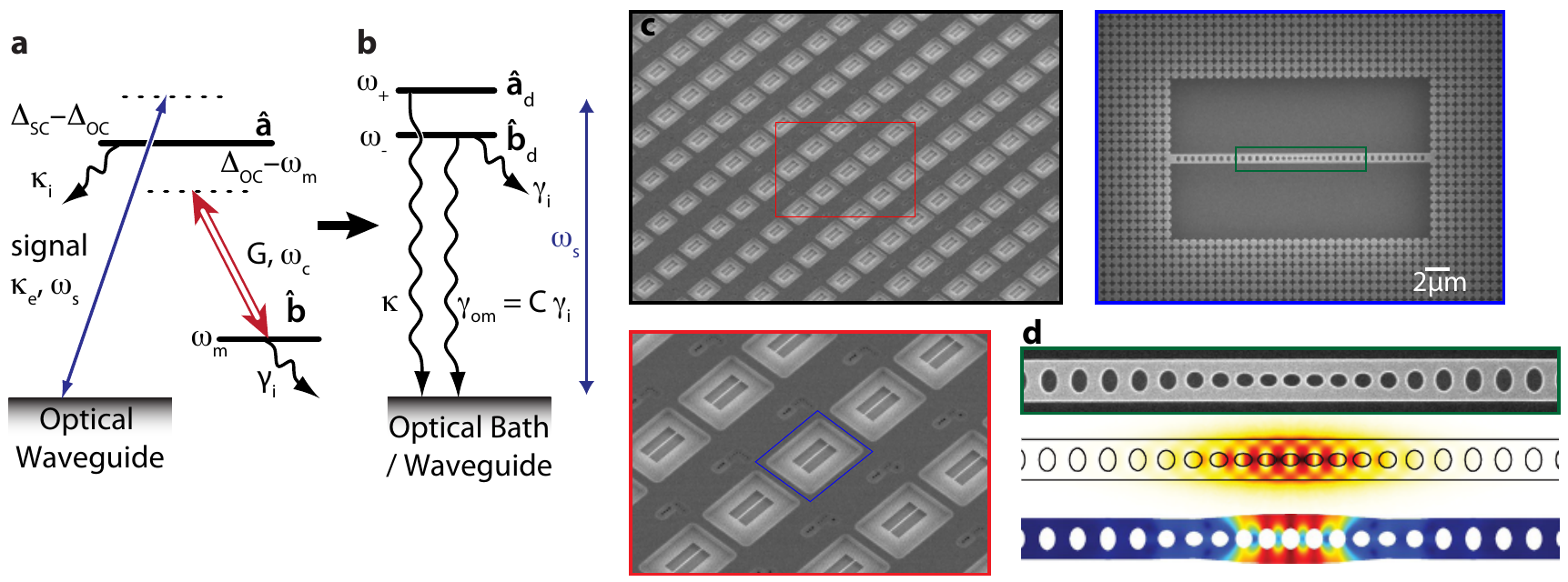}
\end{center}
\caption{\textbf{Optomechanical System.}  \textbf{a,} Level-diagram picture, showing three ``levels'' representing the optical mode $\hat{a}$, mechanical mode $\hat{b}$ and the ``bath'' of optical waveguide modes. \textbf{b,} The control beam at $\omega_c$ drives the transition between the optical and mechanical mode,  dressing the optical and mechanical modes, resulting in the dressed state picture with dressed modes $\hat{a}_d$ and $\hat{b}_d$.  \textbf{c,} Scanning electron micrographs (SEM) of an array of optomechanical crystal nanocavities. \textbf{d,} From top to bottom: scanning electron micrograph (SEM) of a zoomed-in region showing the OMC defect region; FEM simulation results for the optical field showing the electrical field intensity $|\m E(\m r)|$; FEM simulated mechanical mode with the total displacement $|\m Q(\m r)|$ shown. 
\label{fig1}}
\end{figure*}

In the solid state, EIT has been demonstrated in quantum wells, dots, and N-V centers~\cite{Phillips2003,Santori2006,Xu2008}. The fast dephasing rates and inhomogeneous broadening of solid state electronic resonances, however, has led to a plethora of other methods and techniques. Elegant experiments with stimulated Brillouin scattering (SBS) in fibers~\cite{Thevenaz2008}, and coherent population oscillations (CPO)~\cite{Bigelow2003} have been used to delay intense classical light. Alternatively, for quantum storage and buffering, techniques based on photon-echo spectroscopy (e.g. CRIB~\cite{Afzelius2009} and AFC~\cite{Riedmatten2008}) have been used successfully to achieve solid-state quantum memories. In chip-scale photonics, proposals for dynamically tunable arrays of cavities, displaying EIT, are an intriguing  analogy to ensembles of atoms and provide a route to slowing and stopping light all-optically~\cite{yanik04}. Generally, the elements in the arrays have consisted of coupled optical or plasmonic resonances, and have been demonstrated with couplings engineered to give rise to Fano-like interference~\cite{Xu2006}.  A significant limitation in these all-photonic systems, however, is the short optical resonance lifetime.  With optomechanics, EIT arising from optically controlled interactions between engineered optical and mechanical modes provides the means for tunable delays on the order of the mechanical resonance lifetime, which due to the lower mechanical frequency, can be orders of magnitude longer than the lifetime of the optical mode. Such delays are attainable at any wavelength and in any material which is of high optical and mechanical quality (indeed, recent circuit cavity electromechanics experiments in the strong-coupling regime have demonstrated group velocity control at microwave frequencies~\cite{Teufel2010a}). Additionally, the on-chip nature and Si compatibility of many optomechanical systems~\cite{Li2008} suggest that arrays of such structures may be possible~\cite{Notomi08}, allowing for the dynamic slowing and storage of light pulses~\cite{Chang2010}. 


EIT in optomechanical systems can be understood physically as follows.  The conventional radiation pressure interaction between a near-resonant cavity light field and mechanical motion is modeled by the nonlinear Hamiltonian $H_\text{int} = \hbar g \hat{a}^\dagger \hat{a} (\hat{b}+\hat{b}^\dagger)$, where $\hat{a}$ and $\hat{b}$ are the annihilation operators of photon and phonon resonator quanta, respectively, and $g$ is the optomechanical coupling rate corresponding physically to the shift in the optical mode's frequency due to the zero-point fluctuations of the phonon mode.  By driving the system with an intense red-detuned optical ``control'' beam at frequency $\omegap$, as shown in Fig.~\ref{fig1}a, the form of the effective interaction changes (in the resolved sideband limit) to that of a beam-splitter-like Hamiltonian $H_\text{int} = \hbar G (\hat{a}^\dagger \hat{b} + \hat{a}\hat{b}^\dagger)$.  Here, the zero-point-motion coupling rate $g$ is replaced by a much stronger parametric coupling rate $G=g\sqrt{\npump}$ between light and mechanics, where $\npump$ is the stored intracavity photon number induced by the control beam.  Viewed in a dressed-state picture, with the control beam detuning set to $\DeltaOP\equiv\omega_o-\omegap \cong \omega_m$, the optical and mechanical modes $\hat{a}$ and $\hat{b}$ become coupled (denoted $\hat{a}_d$ and $\hat{b}_d$ in Fig.~\ref{fig1}b). The dressed mechanical mode, now effectively a phonon-photon polariton, takes on a weakly photonic nature, coupling it to the optical loss channels at a rate $\gamma_\text{om} \equiv C \gamma_i$, where the optomechanical cooperativity is defined as $C \equiv 4G^2/\kappa\gamma_i$ for an optical cavity decay rate of $\kappa$.  

The drive-dependent loss rate $\gamma_\text{om}$ has been viewed in most previous studies as an incoherent, quantum-limited loss channel, and was used in recent experiments to cool the mechanical resonator close to its quantum ground state~\cite{Rocheleau2010}. In the dressed mode picture, in analogy to the dressed state view of EIT~\cite{fleischhauer05}, it becomes clear that a coherent cancellation of the loss channels in the dressed optical and mechanical modes is possible, and can be used to switch the system from absorptive to transmittive in a narrowband around cavity resonance.  Much as in atomic EIT, this effect causes an extremely steep dispersion for the transmitted probe photons, with a group delay on resonance of (see Appendix A)
\be
\tau^\textrm{(T)}|_{\omega = \omega_m} = \frac{2}{\gamma_i} \frac{(\kappa_e/\kappa)C}{(1+C)(1- (\kappa_e/\kappa) + C)},
\ee

\noindent where $\kappa_{e}$ is the optical coupling rate between the external optical waveguide and the optical cavity, and the delay is dynamically tunable via the control beam intensity through $C$.

Nano- and micro-optomechanical resonators take a variety of forms, among which optomechanical crystals (OMC) have been used to demonstrate large radiation-pressure-induced interaction strengths between gigahertz mechanical and near-infrared optical resonances~\cite{eichenfield09}. The nanobeam OMC cavity used in this study (Figs.~\ref{fig1}c and ~\ref{fig1}d) utilizes a periodic free-standing Si structure to create high-$Q$ co-localized optical and mechanical resonances. These devices can be printed and etched into the surface of a Si chip in large arrays (Fig.~\ref{fig1}c), and are designed to operate optically in the telecom band ($\lambda_o =1550~\text{nm}$) and acoustically at microwave frequencies ($\omega_m/2\pi = 3.75~\text{GHz}$). The theoretical optomechanical coupling rate $g$ between co-localized photon and phonon modes is $g/2\pi \approx 800~\text{kHz}$.  By optimizing both the defect and crystal structure, an intrinsic optical decay rate of $\kappa_i/2\pi \approx 290~\text{MHz}$ is obtained for the optical mode, placing the optomechanical system in the resolved sideband regime ($\omega_m/\kappa_i \gg 1$) necessary for EIT.  The corresponding mechanical resonance is measured to have an intrinsic damping rate of $\gamma_i/2\pi \approx 250~\text{kHz}$ ($T=8.7$~K).  Light is coupled into and out of the device using a specially prepared optical fiber taper, which when placed in the near-field of the nanobeam cavity couples the guided modes of the taper evanescently to the optical resonances of the nanobeam (see Appendix A for details of the optical cavity loading).

\begin{figure*}[ht!]
\begin{center}
\includegraphics[width=17.5cm]{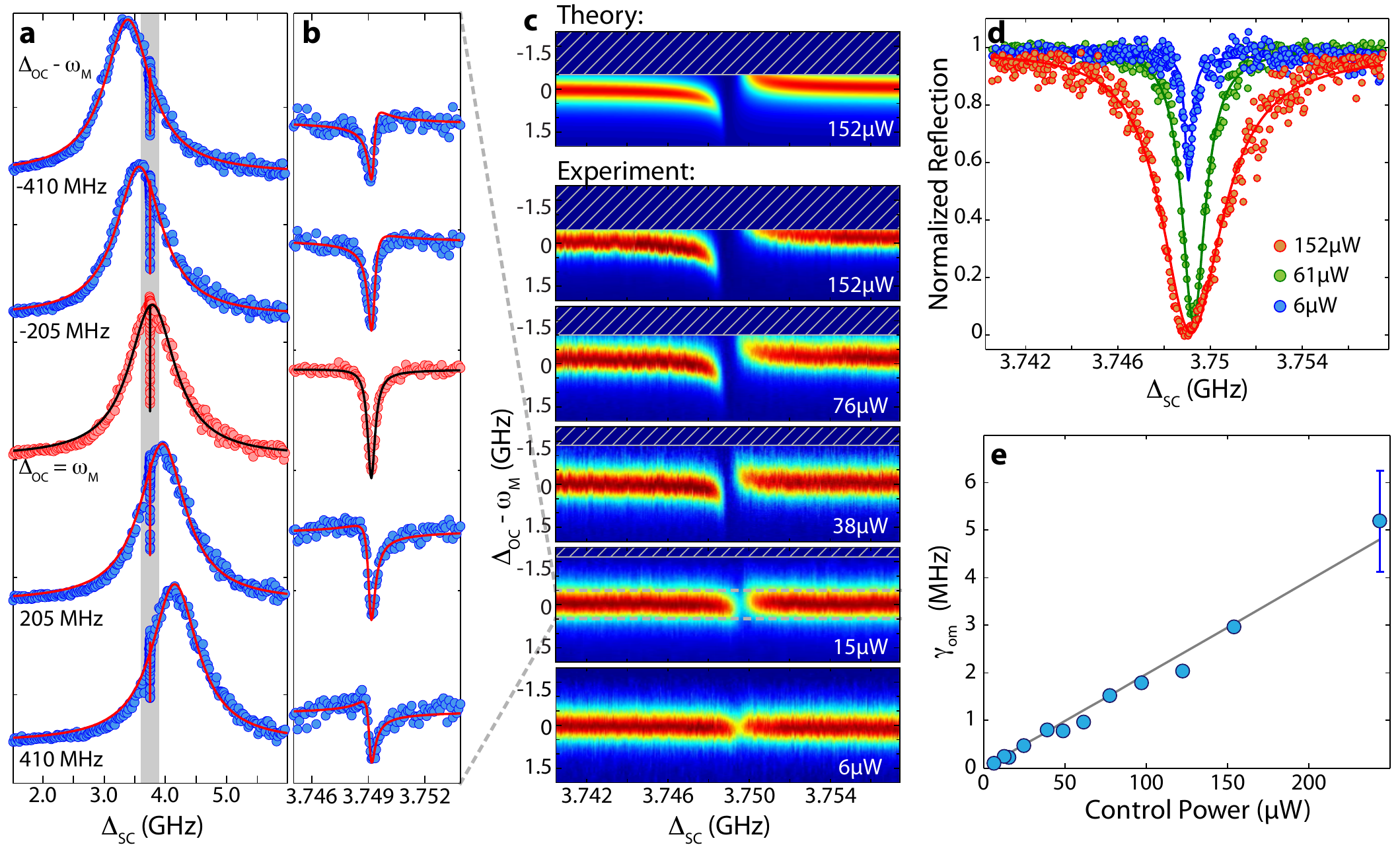}
\end{center}
\caption{\textbf{Optical Reflection Response ($T=8.7$~K).} \textbf{a,} Measured normalized reflection (dots) of the signal beam as a function of the two-photon detuning for a control beam power of $15$~$\mu$W.  \textbf{b,} Zoom-in of the reflected signal about the transparency window. Each spectrum in \textbf{a} and \textbf{b} corresponds to a different control laser detuning ($\DeltaOP-\omega_{m}$) as indicated.  Solid curves correspond to model fits to the data (see Appendix A). \textbf{c}, Intensity plots for the signal beam reflection as a function of both control laser detuning ($\DeltaOP$) and two-photon detuning ($\tpd$) for a series of different control beam powers (as indicated). The hatched areas are unstable regions for the control laser detuning at the given input power. The top plot is the theoretically predicted reflection spectrum for the highest control beam power. \textbf{d,} Transparency window versus control beam power for control laser detuning $\DeltaOP\approx\omega_{m}$. \textbf{e,} Transparency window bandwidth ($\gamma_\text{om} = 4G^2/\kappa$) versus control beam power. The solid line represents the fit to the model (see Appendix A) which determines $g$.
\label{fig2}}
\end{figure*}

\begin{figure}[ht!]
\begin{center}
\includegraphics[width=14cm]{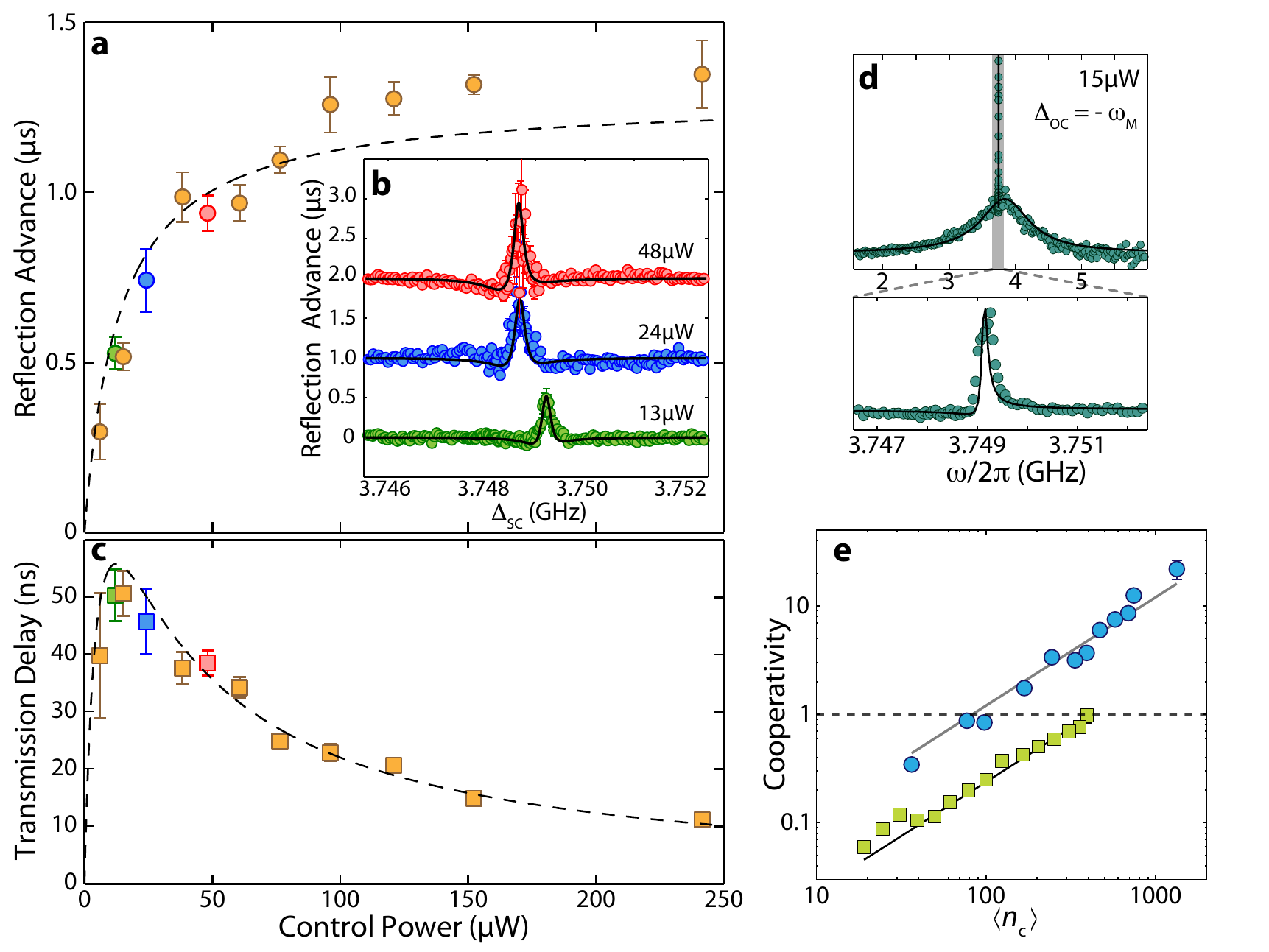}
\end{center}
\caption{\textbf{Measured Temporal Shifts and Amplification.} \textbf{a,} Maximum measured reflected signal advance as a function of the control beam power. \textbf{b,} Measured reflected signal advance versus two photon detuning, $\tpd$. Solid curves correspond to fit from model (see Appendix A).  Curves at different control powers are shifted for clarity. \textbf{c,} Inferred maximum transmitted signal delay versus conttol beam power.  Dashed lines in \textbf{a} and \textbf{c} are theoretical advance/delay times predicted from model of optomechanical system based upon intensity response of the optomechanical system. \textbf{d,} Measured signal reflection signal as a function of two photon detuning for the control laser blue detuned from the cavity. \textbf{e,} Measured cooperativity for sample temperature of $296$~K ($\square$) and $8.7$~K ($\circ$) as a function of the average number of control photons inside the cavity. \label{fig3}}
\end{figure}

In order to characterize the near-resonance optical reflection of the cavity system, a sideband of the control beam is created using electro-optic modulation (see Methods and Appendix C), forming a weak signal beam with tunable frequency $\omega_s$. The results of measurements performed at a cryogenic temperature of $8.7~\text{K}$ are shown in Fig.~\ref{fig2}. Here, the control beam laser power was varied from $6~\mu\text{W}$ ($\npump = 25$) to nearly $250~\mu\text{W}$ ($\npump = 1040$). The frequencies of both the control and signal beams are swept in order to map out the system dependence upon control-cavity detuning, $\DeltaOP$, and the two-photon detuning, $\tpd = \omega_s - \omegap$. The resulting reflected optical signal intensity, separated from the control beam via a modulation and lock-in technique (see Methods and Appendix C), is shown in Fig.~\ref{fig2}(a) for a series of control laser detunings.  Visible in each of the plots is a broad resonance corresponding to the bare optical cavity response with loaded linewidth $\kappa/2\pi \approx 900~\text{MHz}$.  A much narrower reflection dip feature, corresponding to the transparency window, can also be seen near the cavity line center.  The position of the narrow reflection dip tracks with a two-photon detuning equal to the mechanical resonance frequency, $\tpd\approx \omega_m$. This region is shown in more detail in Fig.~\ref{fig2}b, where the Fano-like structure of the optical response is apparent. Each curve in Figs.~\ref{fig2}a and ~\ref{fig2}b is a horizontal slice of the data presented in Fig~\ref{fig2}c, where the reflectivity is plotted as a function of both $\omegap$ and $\tpd$. The transparency window is shown to be fully controllable via the appled light field, the window expanding and contracting with the control beam laser power (Fig.~\ref{fig2}d). At the maximum stable control power (unstable regions due to a thermo-optic bistability induced by optical absorption are shown as hatched regions in Fig.~\ref{fig2}c), a transparency window approaching $5~\text{MHz}$ is obtained. 

A model fit to the reflection spectra (see Appendix A) are shown as solid curves in Figs.~\ref{fig2}a and ~\ref{fig2}b. The resulting fit values for $\gamma_\text{om}=4G^2/\kappa$ for each control power are shown in Fig.~\ref{fig2}e. A linear fit to the extracted data yields a value for the zero-point-motion coupling constant of $g/2\pi = 800~\text{kHz}$, in agreement with the value obtained from independent optical transduction measurements of the thermal Brownian motion of the mechanical oscillator~\cite{eichenfield09}.  In addition to the intensity response of the optomechanical cavity there is the phase response, which provides a measure of the group delay of the modulated optical signal beam as it passes through the cavity.  For the $89~\text{kHz}$ modulation of the signal beam used in our experiments, corresponding to a free-space signal wavelength of $\sim3.4~\text{km}$, phase shifts between the modulation sidebands and the signal carrier on the order of a fraction of a radian are measured in the region where $\tpd$ is within a mechanical linewidth of $\omega_m$. The measured phase-shifts for the reflected signal beam correspond to advances in time of the modulated signal, pointing to causality-preserving superluminal effects.  A plot of the peak effective signal advance versus control beam power is plotted in Fig.~\ref{fig3}a, ascertained from a fit to the reflection phase response spectra (Fig.~\ref{fig3}b).  For the highest control power, the reflected signal is advanced by $1.3~\mu s$, roughly 7000 times longer than the bare optical cavity lifetime.

The delay in transmission is directly related to the advance upon reflection through the bare cavity transmission contrast (measured independently; see Appendices A and C). As such, we plot the corresponding transmission group delay of the signal in Fig.~\ref{fig3}c.  The theoretical delay/advance of the modulated signal beam for system parameters given by fits to the EIT intensity spectra are shown as dashed curves in Figs.~\ref{fig3}a and  \ref{fig3}c, indicating good agreement with the measured phase response.  As can be seen in this data, the maximum measured transmission delay is $\tau^\textrm{(T)}\approx 50$~ns, which although corresponds to significant slowing of light to a velocity of $v_{g}\approx 40$~m/s through the few micron long structure, is much smaller than the measured reflected signal advance or the limit set by the intrinsic mechanical damping ($2/\gamma_{i}\approx 1.4$~$\mu$s).  This is due to the weak loading of the optical cavity in these experiments (see Appendix A), and the resulting small fraction of transmitted light that actually passes through the cavity.

In addition to the observed EIT-like behavior of the optomechanical system, a similar phenomena to that of Electromagentic Induced Absorption (EIA)~\cite{Lezama1999} in atomic systems can be realized by setting the detuning of the control beam to the blue side of the optomechanical cavity resonance ($\DeltaOP<0$).  Under blue-detuned pumping, the effective Hamiltonian for the optical signal and mechanical phonon mode becomes one of parametric amplification, $H_\text{int} = \hbar G (\hat{a}^\dagger \hat{b}^\dagger + \hat{a}\hat{b})$.  The measured reflection spectrum from the OMC is shown in Fig.~\ref{fig3}d, where the reflectivity of the cavity system is seen to be enhanced around the two-photon detuning $\tpd \sim \omega_m$, a result of the increased ``absorption'' (feeding) of photons into the cavity. As discussed further in the Appendix A, at even higher control beam powers such that $C \gtrsim 1$, the system switches from EIA to parametric amplification, resulting in optical signal amplification, and eventually phonon-lasing.

Reflection spectroscopy at room temperature ($296$~K) of the optomechanical cavity has also been performed (presented in the Appendix D), and yields similar results to that of the cryogenic measurements, albeit with a larger value of $\npump$ required to reach a given cooperativity (see Fig.~\ref{fig3}e) due to the larger intrinsic mechanical dissipation at room temperature ($\gamma_i=2\pi\times1.9~\text{MHz}$).  Beyond the initial demonstrations of EIT and EIA behaviour in the OMC cavities presented here, it is fruitful to consider the bandwidth and signal delay limits that might be attainable with future improvements in device material or geometry.  For instance, the transparency bandwidth of the current devices is limited by two-photon absorption of the control beam in the silicon cavities; a move to larger bandgap dielectric materials, such as silicon nitride, should allow intra-cavity photon numbers of $10^6$ (limited by linear material absorption), resulting in a transparency window approaching $G=g\sqrt{\npump}\sim 2\pi$(1~GHz).  Also, recent research into low-loss GHz mechanic resonators~\cite{Nguyen07} should enable slow light optical delays approaching $10~\mu$s at room temperature, roughly a path length of a kilometer of optical fiber.  Much like the acoustic wave devices used in electronic systems~\cite{Lakin1995}, optomechanical devices with these attributes would enable chip-scale microwave photonic systems capable of advanced signal processing in the optical domain, such as that needed for emerging broadband wireless access networks or more specialized applications such as true-time delays in radar systems~\cite{Boyd2009}.           

The limiting factor for quantum applications of optomechanical systems is the re-thermalization time of the mechanical resonator, $\tau_\text{th}=\hbar Q_m/k T$, which in the case of a quantum optical memory represents the average storage time of a single photon before excitation of the system by a thermal bath phonon.  For the devices studied here, despite the small optically-cooled phonon occupancy of the resonator ($\sim 2.1$ phonons from the measured cooperativity), the re-thermalization time is only $\tau_\text{th}\approx 12$~ns at the measurement temperature of $8.7$~K ($50$~ps at room temperature).   Reducing the operating temperature further to a value below $100$~mK (routinely attained in a dilution refigerator), would not only increase the re-thermalization time through a lower bath temperature, but should also result in a significant increase in the mechanical $Q$-factor.  Taken together, the resulting re-thermalization time in the current OMC devices at $T=100$~mK is likely to be on the order of $100$~$\mu$s, which although not nearly as long as what has been achieved in atomic systems~\cite{Zhang2009}, still represents a substantial storage time compared to the realizable GHz bandwidth of the system.  Additionally, optomechanical processes similar to the EIT behavior measured here have also been proposed~\cite{Stannigel2010,Safavi-Naeini2010a} to provide an optical interface between, for instance, atomic and superconducting circuit quantum systems, enabling the formation of hybrid quantum networks.

 \section*{Acknowledgements}

The authors would like to thank Keith Schwab for providing the microwave modulation source used in this work.  This work was supported by the DARPA/MTO ORCHID program through a grant from AFOSR, and the Kavli Nanoscience Institute at Caltech. ASN gratefully acknowledges support from NSERC.

\section*{Methods}
\textbf{Fabrication:}
\small{The nano-beam cavities were fabricated using a Silicon-On-Insulator wafer from SOITEC ($\rho=4$-$20$~$\Omega\cdot$cm, device layer thickness $t=220$~nm, buried-oxide layer thickness $2$~$\mu$m). The cavity geometry is defined by electron beam lithography followed by inductively-coupled-plasma reactive ion etching (ICP-RIE) to transfer the pattern through the $220~\text{nm}$ silicon device layer. The cavities were then undercut using $\text{HF:H}_2\text{O}$ solution to remove the buried oxide layer, and cleaned using a piranha/HF cycle. The dimensions and design of the nano-beam will be discussed in detail elsewhere.}

\textbf{Experimental Set-up: }\small{We demonstrate EIT via reflection measurements of the optically pumped system at varying $\npump$. Using the experimental setup shown in Appendix C, a laser beam at $\omegap$ (the control beam) is sent through an electro-optical modulator (EOM) with drive frequency $\tpd$, creating an optical sideband at frequency $\omega_{s}$ (the signal beam), which is amplitude modulated at $\omega_\text{LI}/2\pi = 89~\text{kHz}$. Since the control beam is detuned from the cavity by $|\DeltaOP|\gg\kappa$, it is effectively filtered when looking in reflection, while the modulated signal beam at $\omegap\pm \tpd$ (where the sign is that of $\DeltaOP$), is near-resonance with the optical cavity and is reflected.  The reflected signal beam is detected using a 12 GHz New Focus PIN photo-diode, with the output electrical signal sent to a lock-in amplifier where the component related to the modulated tone ($\omega_\text{LI}=89$~kHz) is amplified and sent to an oscilloscope. By scanning both the laser frequency $\omegap$ and the two-photon detuning $\tpd$, a full map of the reflectivity is obtained. Additionally, by using a lock-in amplifier, the phase of the modulated signal sidebands relative to the carrier can be measured, giving a direct measurement of the group delay imparted on the optical signal beam by the optomechanical cavity.}

\bibliographystyle{naturemag}

\appendix

\section{Theory of Optomechanical EIT, EIA and Parametric Amplification}
Here we provide a theoretical treatment of some of the main aspects of EIT~\cite{agarwal10,Weis2010,fleischhauer05,lukin03}, EIA~\cite{Lezama1999} and parametric amplification~\cite{Louisell1961,Yariv1965,Mollow1967}  in
optomechanical systems. Modeling the
optomechanical system with the Hamiltonian
\be \op{H}{} = \hbar
\omega_o \opdagger{a}{}\op{a}{} + \hbar \omega_m
\opdagger{b}{}\op{b}{}+\hbar g (\opdagger{b}{} + \op{b}{})
\opdagger{a}{} \op{a}{} + i\hbar \sqrt{\frac{\kappae}{2}} \alpha_{\text{in},0} e^{-i \omegap t}(\op{a}{} - \opdagger{a}{}) , \ee
it is possible to linearize the operation of the system, under the influence of a control laser
at $\omegap$, about a particular steady-state given by intracavity photon amplitude $\alpha_0$ and a static phonon shift $\beta_0$. The interaction of the mechanics and pump photons at $\omegap$ with secondary ``probe'' photons at $\omega_s = \omegap\pm\tpd$ with  two-photon detuning $\tpd$ can then be modeled by making the substitutions
\be
\op{a}{} \rightarrow \alpha_0 e^{-i\omegap t}  + (\alpha_- e^{-i(\omegap + \tpd) t} + \alpha_+e^{-i(\omegap - \tpd) t}), \qquad
\op{b}{} \rightarrow \beta_0 + \beta_- e^{-i\tpd t}.
\ee
Assuming that the pump is much larger than the probe, $|\alpha_0| \gg |\alpha_\pm|$,  the pump amplitude
is left unaffected and the equations for each sideband amplitude $\alpha_\pm$ are found to be
\bea
\pm i \omega \alpha_\pm &=& -\left(i \DeltaOP + \frac{\kappa}{2}\right) \alpha_\pm - i g \alpha_0 \beta_\pm - \sqrt{\frac{\kappae}{2}}\alpha_{\text{in},\pm}, \label{eqn:alphapmsb} \\
- i \omega \beta_- &=& -\left( i \omega_m + \frac{\gamma_i}{2}\right) \beta_- - i g(\alpha_0^\ast \alpha_- + \alpha_0 \alpha_+^\ast) - \sqrt{\gamma_i} \beta_{\textrm{in},-} \label{eqn:betamsb}.
\eea
We have defined  $\DeltaOP = \omega^\prime_o - \omegap$ as the pump detuning from the optical cavity (including the static optomechanical shift, $\omega^\prime_o$),  and $\beta_+  = \beta_-^\ast$. In these situations it is typical to define $G = g \alpha_0$, as the effective optomechanical coupling rate between a sideband and the mechanical subsystem, mediated by the pump.

\subsection{Red-detuned pump: Electromagnetically Induced Transparency}
\begin{figure}[ht!]
\begin{center}
\includegraphics[width=14cm]{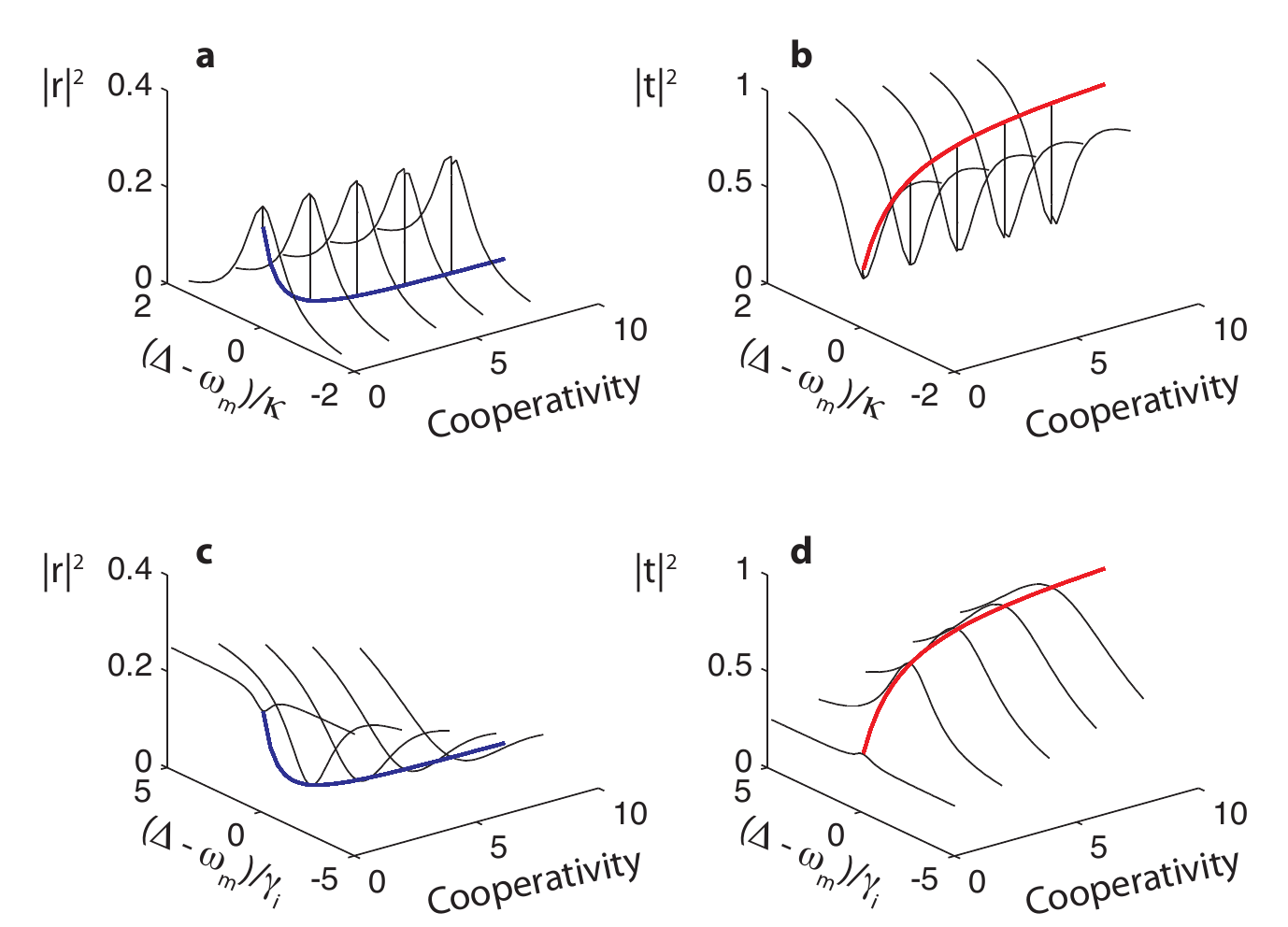}
\end{center}
\caption{\textbf{Electromagnetically Induced Transparency Spectra. a, c,} The reflected signal amplitude, as a function of two-photon detuning $\tpd$ for the case where $\DeltaOP = \omega_m$. In \textbf{b, d,} the corresponding plots for transmission are shown. The broadening of the transmission window, and the saturation of the transmission peak, and reflection dip are evident in \textbf{c, d} respectively.
\label{fig:EIT}}
\end{figure}
With the pump detuned from the cavity by a two-photon detuning $\tpd$, the spectral
selectivity of the optical cavity causes the sideband populations to be skewed in
a drastic fashion. It is then an acceptable approximation to neglect one of these
sidebands, depending on whether the pump is on the red or blue side of the
cavity. When the pump resides on the red side ($\DeltaOP > 0$), the $\alpha_+$ is
reduced and can be neglected. This is the rotating wave approximation (RWA)
and is valid so long as $\tpd \gg \kappa$.

\begin{figure}[ht!]
\begin{center}
\includegraphics[width=14cm]{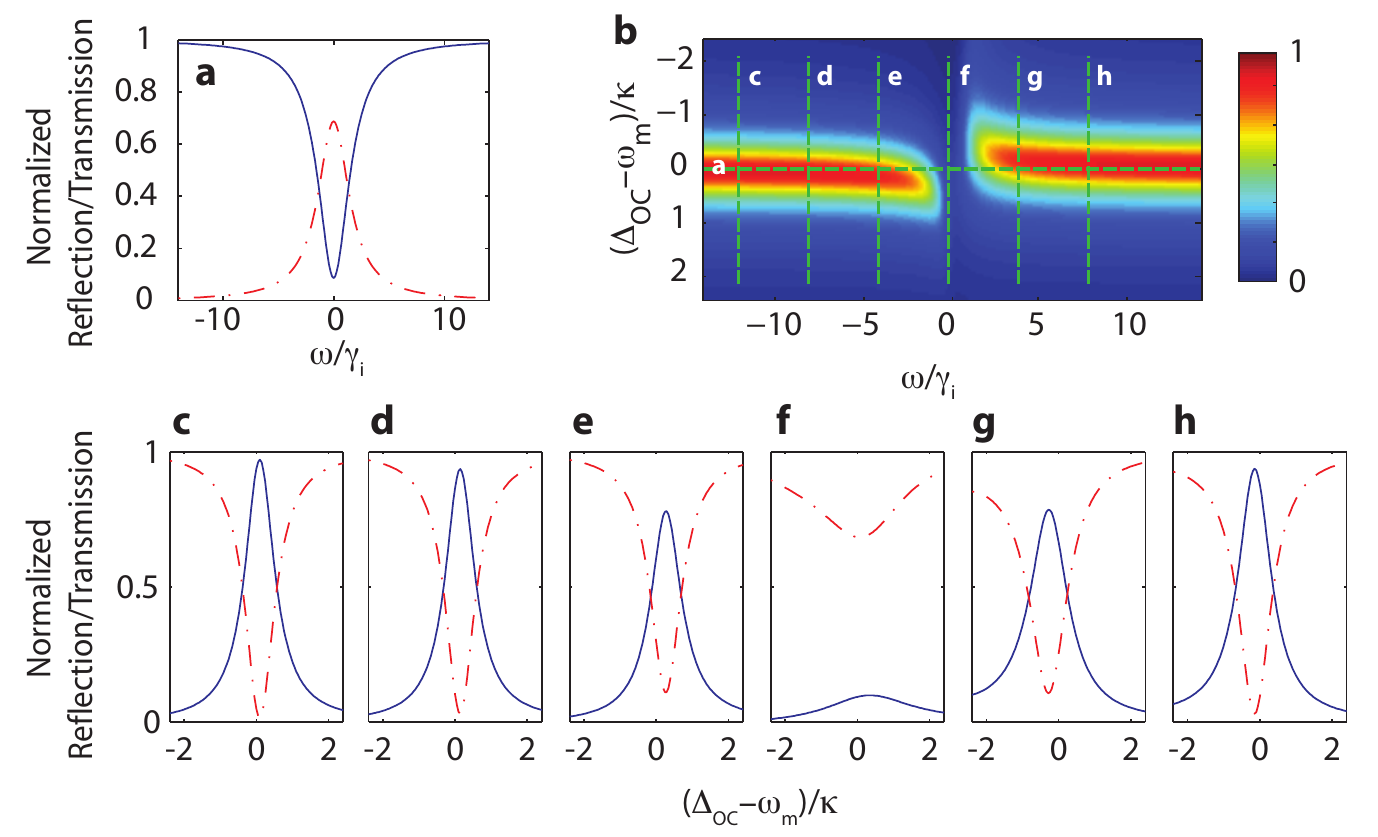}
\end{center}
\caption{\textbf{a}, normalized reflection (solid blue line) and transmission (dot-dash red line) signal for $\DeltaOP=\omega_m$ versus the normalized two-photon detuning frequency ($\tpd/\gamma_i$). \textbf{b} normalized reflection signal map as a function of the normalized pump detuning ($(\DeltaOP-\omega_m)/\kappa$) and the normalized two-photon detuning frequency. Each dash line corresponds to the curves shown on textbf{a} and \textbf{c-h}. \textbf{c-h} normalized reflection (solid blue line) and transmission (dot-dash red line) signals as a function of normalized pump detuning. Each curve corresponds to a specific two-photon detuning. For \textbf{f}, $\tpd\approx\omega_m$, then the reflected signal is pratically zero on the vicinity of the resonance condition $\DeltaOP =\omega_m$. The curves where generated based upon the Eq.~\ref{eq:r_omega_red}.
\label{fig:refltrans}}
\end{figure}

Then Eqs. (\ref{eqn:alphapmsb}-\ref{eqn:betamsb}) may be solved for the
reflection and transmission coefficients $r(\omega_s)$ and $t(\omega_s)$ of the side-coupled
cavity system. We find that
\bea \label{eq:r_omega_red}
r(\tpd) &=& -\frac{\kappae/2}{i(\DeltaOP -\tpd) + \kappa/2 + \frac{|G|^2}{i(\omega_m-\tpd)+\gamma/2}}\\
t(\tpd) &=& 1-\frac{\kappae/2}{i(\DeltaOP -\tpd) + \kappa/2 + \frac{|G|^2}{i(\omega_m-\tpd)+\gamma/2}}.
\eea
These equations are plotted in Figs.~\ref{fig:EIT} and \ref{fig:refltrans}.

\subsubsection{Group Delay}

For the red-detuned system, the existence of an effective transparency on transmission
makes the group delay imparted on the pulse an interesting quantity. To calculate the reflection and transmission group delays
we consider a pulse
\be
f(t_o) = \int_0^\infty f(\omega)e^{-i\omega t_o} \textrm{d}\omega,
\ee
where most of the spectrum is confined to a small window ($< 4G^2/\kappa$) about a central signal frequency $\omega_s$. Then the transmitted signal $f^\textrm{(T)}(t_o)$ may be written as
\bea
f^\textrm{(T)}(t_o) &=& \int_0^\infty t(\omega) f(\omega)e^{-i\omega t_o} \textrm{d}\omega\nonumber\\
&=& e^{-i\omega_s t_o}  \int_{-\infty}^\infty t(\omega_s + \delta) f(\omega_s + \delta)e^{-i\delta t_o} \textrm{d}\delta\nonumber\\
&=&  e^{-i\omega_s t_o}  \int_{-\infty}^\infty t(\omega_s)\left(1 + \frac{1}{t(\omega_s)}\left. \frac{\textrm{d}t}{\textrm{d}\omega}\right|_{\omega_s}\delta+o(\delta^2)\right) f(\omega_s + \delta)e^{-i\delta t_o} \textrm{d}\delta\nonumber\\
&\approx& e^{-i\omega_s t_o}  \int_{-\infty}^\infty t(\omega_s)f(\omega_s + \delta)e^{-i\delta (t_o-\tau^\textrm{(T)})} \textrm{d}\delta\nonumber\\
\eea
The last line implies that $f^\textrm{(T)}(t_o) \approx f(t_o-\tau^\textrm{(T)})$, where
\be
\tau^\textrm{(T)} = \mathcal{R}\left\{\frac{-i}{t(\omega_s)}\frac{\textrm{d}t}{\textrm{d}\omega}\right\}.
\ee
The reflection group delay may also be defined analogously,
\be
\tau^\textrm{(R)} = \mathcal{R}\left\{\frac{-i}{r(\omega_s)}\frac{\textrm{d}r}{\textrm{d}\omega}\right\}.
\ee
With the signal sent at a two-photon detuning $\tpd = \omega_m$, we find
\be\label{eq:tau_T_red}
\tau^\textrm{(T)}|_{\tpd = \omega_m} = \frac{2}{\gamma_i} \frac{ (\kappa_e/\kappa) C}{(1+C)(1- (\kappa_e/\kappa) + C)} ,
\ee
where the cooperativity $C=4G^2/\kappa\gamma_i$ is a measure of the coupling between the mechanical oscillator and the optical bath.
Under the same conditions we find that group delay for reflection is given by
\be
\tau^\textrm{(R)}|_{\tpd = \omega_m} = -\frac{2}{\gamma_i} \frac{C}{1+C},
\ee
resulting in the limit $C \gg 1$
\be
\tau^\textrm{(T)}|_{\tpd = \omega_m}\rightarrow \frac{2}{\gamma_i}\frac{\kappa_e}{\kappa}\frac{1}{C} \qquad\textrm{and}\qquad \tau^\textrm{(R)}|_{\tpd = \omega_m} \rightarrow- \frac{2}{\gamma_i}.
\ee
A quantity of interest, the delay-bandwidth product can be calculated for the transmitted signal, by taking the product of the signal delay $|\tau^\textrm{(T)}|_\textrm{max}$, and the bandwidth $\Delta \omega = \gamma_i C$, to give us $\Delta \omega\cdot t_d = 2({\kappa_e}/{\kappa})$.

\begin{figure}[ht!]
\begin{center}
\includegraphics[width=17.5cm]{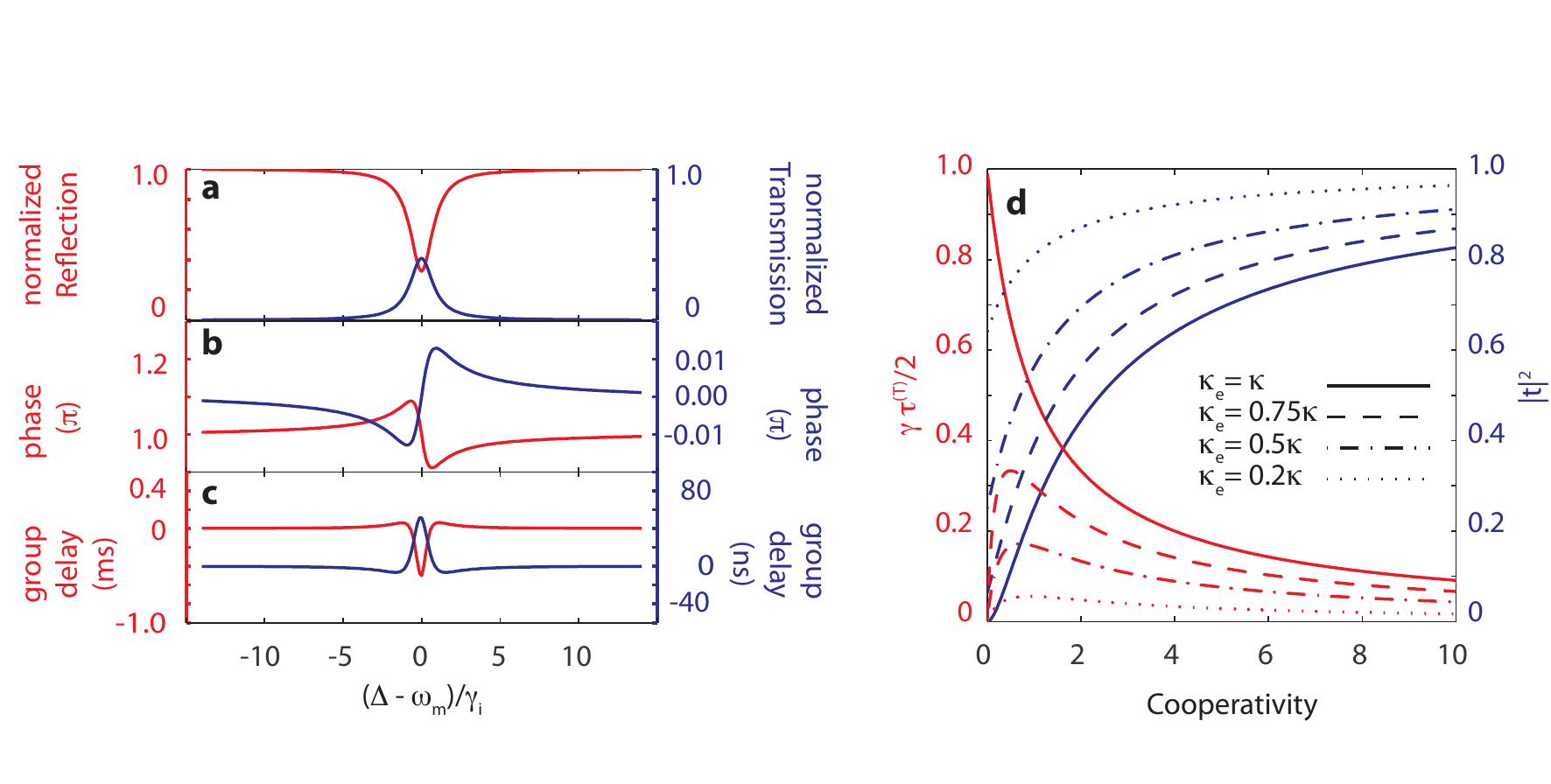}
\end{center}
\caption{\textbf{Phase and Group Delay in EIT.} \textbf{a,} Normalized reflection (red) and transmission (blue) signals, with the phase \textbf{b}, and group delay \textbf{c}, for typical system parameters. \textbf{d,} Maximum delay ($\tau^\text{(T)}_\text{max}$), in units of $\gamma_i/2$, and transmission coefficient ($|t_\text{max}|^2$) for the transmitted signal as a function of the cooperativity for different cavity-waveguide couplings. 
\label{fig:delay_vs_C_trans}}
\end{figure}

Using equations \ref{eq:tau_T_red} and \ref{eq:r_omega_red} we can estimate the maximum delay for our system. The reflection and transmission coefficients at resonance ($\tpd = \DeltaOP = \omega_m)$) are given by
\be
r_\text{max} = -\frac{(\kappa_e/\kappa)}{1+C} \qquad\text{and}\qquad t_\text{max} = \frac{1-(\kappa_e/\kappa)+C}{1+C}.
\ee
For the case where intrinsic optical losses are negligible, i.e. $\kappa_e = \kappa$, the equations for delay and transmission coefficient contrast can be written as
\be
t_\text{max} = \frac{C}{1+C},\qquad
\tau^\textrm{(T)}_\text{max} = \frac{2}{\gamma_i} \frac{1}{(1+C)},
\ee
and are plotted in Fig.~\ref{fig:delay_vs_C_trans}.

\subsection{Blue-detuned pump:  Electromagnetically Induced Absorption and Amplification}

\begin{figure}[ht!]
\begin{center}
\includegraphics[width=14cm]{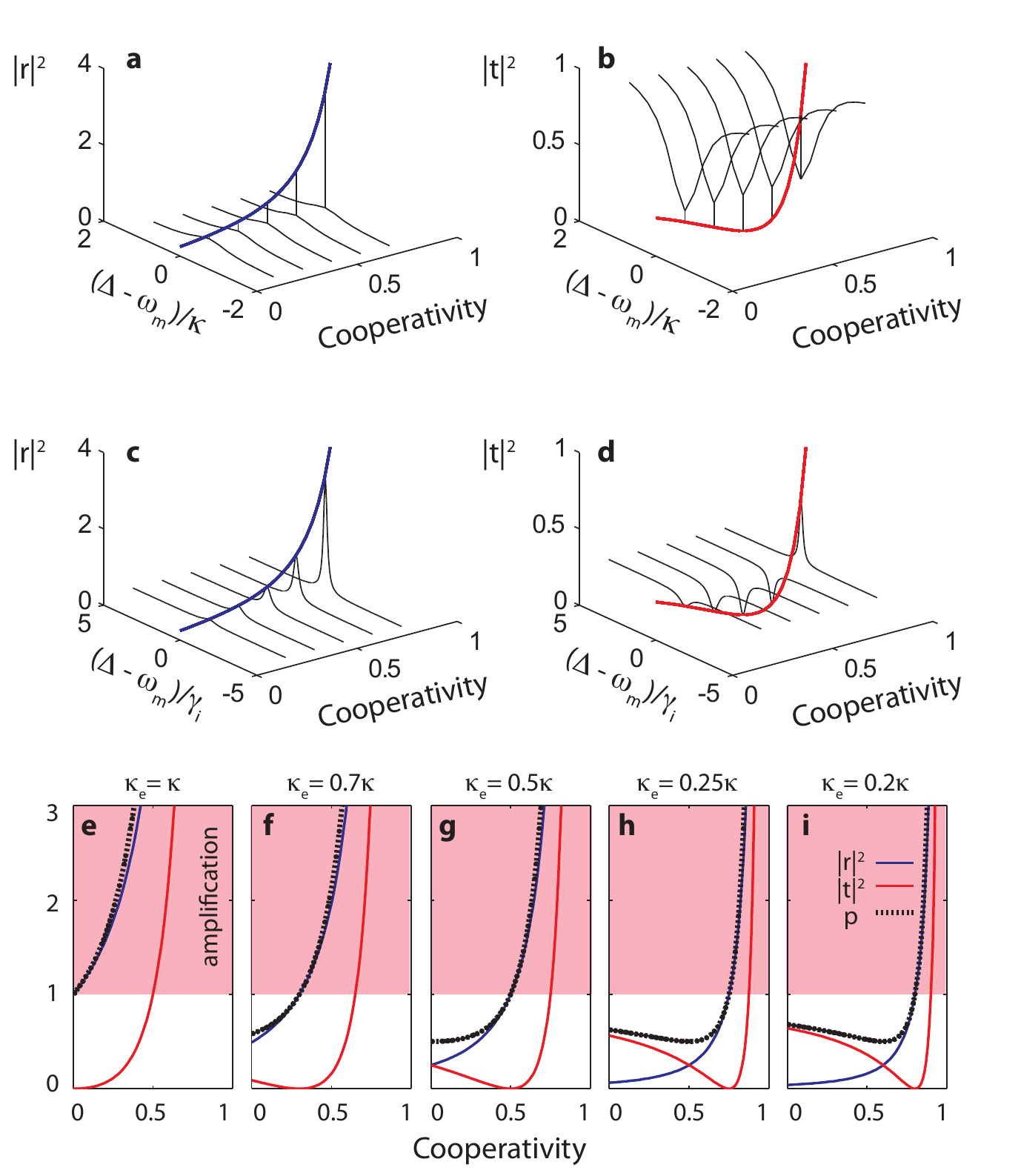}
\end{center}
\caption{\textbf{Electromagnetically Induced Absorption and Amplification Spectra. a, c,} The reflected signal amplitude, as a function of two-photon detuning $\tpd$ for the case where $\DeltaOP = -\omega_m$. In \textbf{b, d,} the corresponding plots for transmission are shown. The increase in the reflected signal is evident in \textbf{a,c}. In \textbf{b,d} we see a reduction in the transmitted signal amplitude, down to zero, and followed by an increase. \textbf{e-i} The amplitude of the reflected (blue), transmitted (red) and total power (dashed black) from the cavity, for various cavity-waveguide coupling efficiencies. The region where $|r|^2, p > 1$ is called the amplification region, and shaded. Note that for $\kappa_e < \kappa/2$, the power from the cavity is at first reduced with higher cooperativity, before increasing and going into the amplification regime at $C > 1 - \kappa_e/\kappa$. This corresponds to electromagnetically induced absorption. 
\label{fig:EIA}}
\end{figure}

By placing the pump at a mechanical frequency away from cavity, on the blue side ($\DeltaOP = -\omega_m$)
we may ignore the $\alpha_-$ sideband of the intracavity photons. The reflection in this case is calculated to be
\be
r_\textrm{A}(\omega_s) = -\frac{\kappae/2}{i(\DeltaOP +\tpd) + \kappa/2 + \frac{|G|^2}{i(\omega_m-\tpd)-\gamma_i/2}}
\label{eqn:rAbluedet}
\ee
and $t_\textrm{A}(\omega_s) = 1 + r_\textrm{A}(\omega_s)$. The linearization which leads to this equation from the full dynamics of the system, only holds below the phonon lasing threshold, $C = 1$, and so we limit ourselves to the case where $C < 1$.
The effective interaction Hamiltonian of the system which is obtained after making the rotating wave approximation to remove terms counter-rotating at the mechanical frequency, is given by~\cite{Aspelmeyer2010}
\be
H_\text{int} = \hbar G (\hat{a}^\dagger \hat{b}^\dagger + \hat{a}\hat{b})\label{eqn:Hamp}.
\ee
This is also the Hamiltonian of a parametric oscillator, whose quantum theory has been known for some time~\cite{Louisell1961,Mollow1967}. The only distinction with our system, is that we consider and measure mainly the reflection and transmission properties of the parametric oscillator, as opposed to its internal dynamics. Using the expression in eqn.~(\ref{eqn:rAbluedet}) we find
\be
r_\text{max} = -\frac{(\kappa_e/\kappa)}{1-C} \qquad\text{and}\qquad t_\text{max} = \frac{1-(\kappa_e/\kappa)-C}{1-C}.
\ee

These expressions are plotted in Fig~\ref{fig:EIA}a-d for a range of cooperativities. The ratio between the power leaving the cavity through the waveguide to the input power is the sum of the reflection and transmission coefficient amplitudes, $p=|t_\text{max}|^2+|r_\text{max}|^2$. As the base-line value, we take $C=0$ for which the emitted power is $p_0 = (\kappa_e/\kappa)^2 + (\kappa_i/\kappa)^2$.

\subsubsection{Weak-Coupling: Electromagnetically Induced Absorption}
At small cooperativities $C \ll 1$ and weak cavity-waveguide coupling $\kappa_e < \kappa_i$, the behaviour of
our system is analogous to what has been observed in atomic gases, and been called Electromagnetically Induced Absorption (EIA)~\cite{Lezama1999}. Under these conditions, $p$ is less than $p_0$, and more of the incoming photons are now absorbed than in the case with $C=0$. As such, the reflection will exhibit an absorption peak, and the transmission an absorption dip. As long as $\kappa_e < \kappa_i$, there will always be a value of $C$ such that absorption is enhanced, as experimentally demonstrated in this paper. In systems where $\kappa_e > \kappa_i$, we find $p > p_0$ for any finite $C$. Plots of the transmission, reflection, and $p$ are shown as a function of cooperativity in Fig.~\ref{fig:EIA}e-i. It can be seen that the slope of $p$ at $C=0$ goes from positive to negative as $\kappa_i$ over-takes $\kappa_e$.  Even with weak cavity-waveguide coupling, at sufficiently high C, the system changes from absorptive to amplifying with $p$ becoming much larger than $p_0$ and 1, as can be seen in the shaded region of Fig.~\ref{fig:EIA}e-i.

\subsubsection{Amplification}

As suggested by the effective Hamiltonian of the system in eqn.~(\ref{eqn:Hamp}), parametric amplification $(p > 1)$ is always possible, and occurs at $C > 1-(\kappa_e/\kappa)$. It is important to note that at perfect coupling, $\kappa_e = \kappa$, amplification will occur for any finite $C$. This can be seen in Fig.~\ref{fig:EIA}, which shows that the transmission is always greater than unity for finite $C$.

\subsubsection{Group Delay}
Following a derivation similar to that in the previous section we arrive at values for maximum group
delay $\tau_A^\textrm{(T)}|_\textrm{max}$, $\tau_A^\textrm{(R)}|_\textrm{max}$ given by
\bea
\tau_A^\textrm{(R)}|_{\tpd = \omega_m} &= &-\frac{2}{\gamma_i} \frac{ C}{1-C},\\
\tau_A^\textrm{(T)}|_{\tpd = \omega_m} &= &\frac{2}{\gamma_i} \frac{\kappae}{\kappa}\frac{ C}{(1-C)(1-\kappae/\kappa -C)}.
\eea
Note that as $C$ increases, and the system switches at $C = 1-\kappa_e/\kappa$ from absorptive to amplifying, and the sign of the group delay on transmission changes from from positive to negative. For systems with strong cavity-waveguide coupling, $\kappa_e \approx \kappa$,  both $\tau_A^\textrm{(R)}|_{\tpd = \omega_m}$ and $\tau_A^\textrm{(T)}|_{\tpd = \omega_m}$ are negative, pointing to causality-preserving superluminal light.

\section{Fabrication }
The nano-beam cavities were fabricated using a Silicon-On-Insulator wafer from SOITEC ($\rho=4$-$20$~$\Omega\cdot$cm, device layer thickness $t=220$~nm, buried-oxide layer thickness $2$~$\mu$m). The cavity geometry is defined by electron beam lithography followed by inductively-coupled-plasma reactive ion etching (ICP-RIE) to transfer the pattern through the $220~\text{nm}$ silicon device layer. The cavities were then undercut using $\text{HF:H}_2\text{O}$ solution to remove the buried oxide layer, and cleaned using a piranha/HF cycle~\cite{Borselli2006}. The dimensions and design of the nano-beam will be discussed in details elsewhere.

\section{Data Measurement and Analysis}
\begin{figure}[ht!]
\begin{center}
\includegraphics[width=17.5cm]{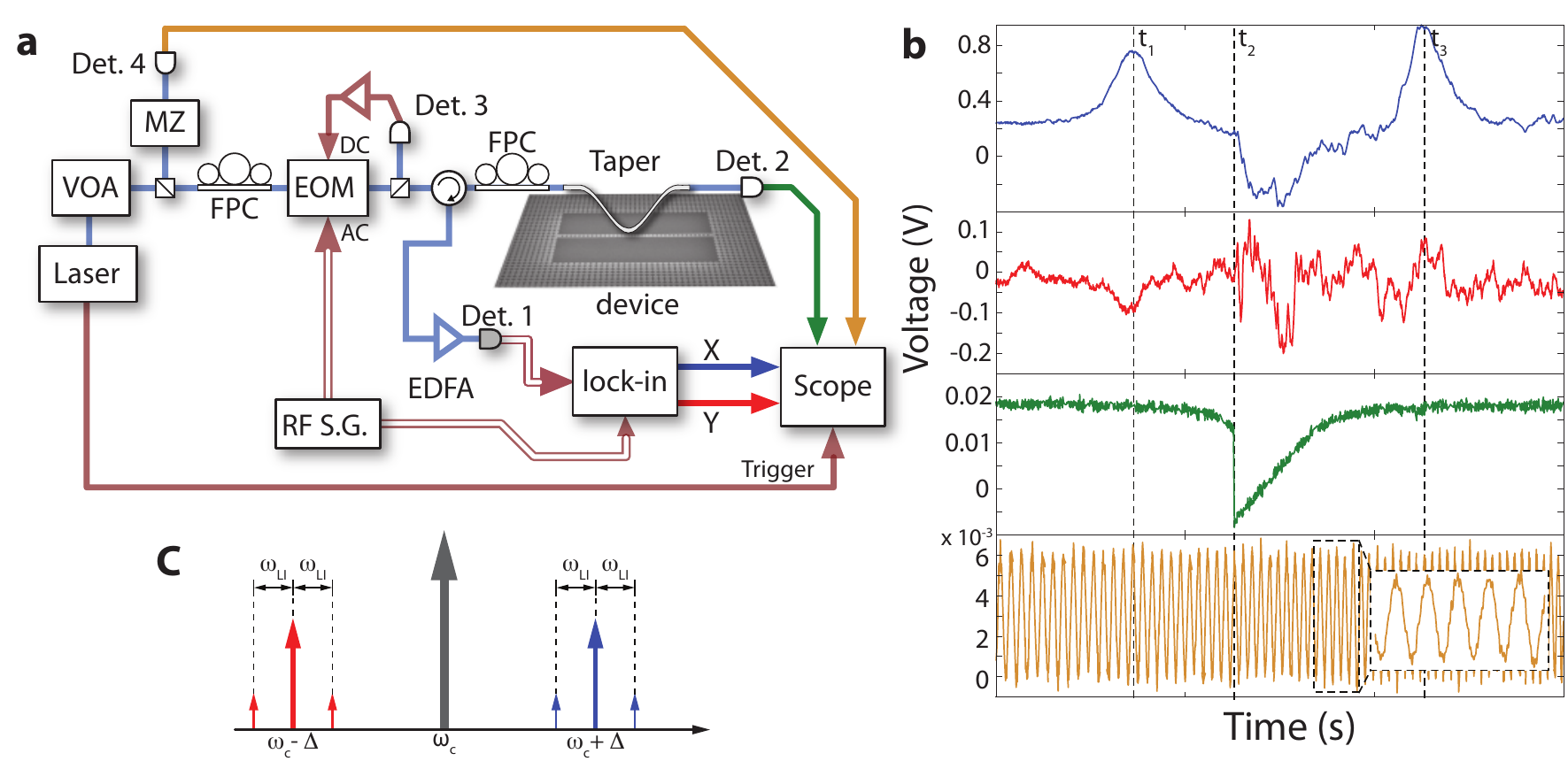}
\end{center}
\caption{\textbf{a}, Experimental setup. \textbf{b}, detected signals for data analysis. From top to bottom, the in phase (blue) and quadrature (red) lock-in signals from the reflected probe side band; the transmitted pump laser (green); and the Mach-Zehnder (MZ) amplitude signal (yellow) for wavelength calibration as explained on the text. \textbf{c}, schematic frequency domain spectra of the pump laser after the EOM. At $t_1$, the probe side band (blue detuned) is at $\omega_s=\omegap+\tpd=\omega_o$, the cavity frequency while at $t_3$ the probe side band is at $\omega_s=\omegap-\omega=\omega_o$.
\label{fig:exp_setup}}
\end{figure}

\subsection{Experimental Setup}
The detailed experimental setup used to measure the EIT window and delay properties of the optomechanical crystal is shown in Fig.~\ref{fig:exp_setup}(a). The setup is designed to record simultaneously the reflected signals from a probe sideband while scanning and recording the pump laser transmission.

As a light source we use a fiber-coupled tunable infrared laser, (New Focus Velocity, model TLB-6328) spanning approximately $60$~nm, centered around $1550$~nm, which has its intensity controlled by a variable optical attenuator (VOA). A small percentage ($10\%$) of the laser intensity is sent to a custom-made fiber coupled Mach-Zehnder (MZ) interferometer and has its intensity detected by a photodetector (Det.4, New Focus Nanosecond Photodetector, model 1623) for further wavelength calibration. To minimize polarization dependent losses on the electro-optical-modulator (EOM), a fiber polarization controller (FPC) is placed before it.

The AC signal used for the electro-optical modulator (EOM) comes from the radio-frequency signal generator (RF S.G., Rohde-Schwarz, SMA-100A). There, a rf-signal carrier at the two-photon detuning frequency ($\tpd/2\pi\sim4$GHz) has its amplitude modulated at the Lock-in detection frequency ($\omega_\text{LI}/2\pi \approx89$~kHz). As a result, the EOM modulation produces two probe sidebands at $\tpd$, where  each of them have a small modulation at the lock-in frequency (see Fig.~\ref{fig:exp_setup}(c)).

A small portion of the signal from the EOM output ($10\%$) is used (Det. 3) as a DC control signal to control for any low frequency power drift during the experiment, by keeping the EOM level locked. The remaining laser light is passed through a circulator, a FPC and then couple to a tapered and dimpled optical fiber (Taper) which has its position controlled with nanometer-scale precision (although vibrations and static electric forces limit the minimum stable spacing between the fiber and device to about 50 nm).

The transmitted light through the taper is detected (Det. 2) and recorded on the oscilloscope (Scope, Agilent, DSO80204B). Any reflected signal coming from the Taper/device is optically amplified by a Erbium-Doped-Fiber-Amplifier (EDFA) and detected by a high-speed photoreceiver (Det. 1, New Focus model, 1554-B) with a maximum transimpedance gain of $1,000$~V/A and a bandwidth ($3$~dB rolloff point) of $12$~GHz. The rf-signal from the photoreceiver is sent to a Lock-in amplifier (L.I., SRS-830). The output from the in-phase and quadrature signals from the L.I. are also recorded on the same oscilloscope. The recorded signal on the oscilloscope is triggered by the sweeping frequency on the pump laser.

Since the pump is detuned from the cavity by $|\DeltaOP|\gg\kappa$, it is filtered on reflection, while the modulated sidebands at $\omegap\pm \tpd$ (where the sign is that of $\DeltaOP$) are reflected and sent to a lock-in amplifier, where the component related to the modulated sideband is amplified and sent to an oscilloscope. Using a lock-in amplifier allows us to measure  the phase-shift on the signal modulated at $\omega_\text{LI}$, giving a direct measurement of the group delay imparted on the signal.

Fig.~\ref{fig:refltrans}(b) shows a sample of the resulting raw data measured on the oscilloscope for the in-phase (blue curve) and out-phase (red curve) reflected signal detected by the lock-in amplifier; the pump transmission spectra (green curve); and the Mach-Zehnder interferometer pattern (yellow curve) used to calibrate the wavelength for the transmission and reflection data. The probe reflection is normalized based upon the transmission contrast for the pump laser at low input power, where the cavity line-shape is not distorted by thermal nonlinearities.

\subsection{Data Analysis}

Here we will show how the amplitude modulation of the signal sideband $\tpd$ is used to measure the reflection ($|r(\omega)|^2$) and delay ($\tau^\text{(R)}$) of the signal reflected from the cavity. The output of the EOM can be written as:
\be
a_\textrm{out}(t) = a_\textrm{in}\left[1+m \left(1+m_\text{LI}\cos(\omega_\textrm{LI}t)\right)\cos(\tpd t)\right],
\ee
where the input field amplitude $a_\textrm{in}(t) = a_o\cos(\tpd t)$, $a_o=\sqrt{P_\text{in}/\hbar\omegap}$, $m$ is the EOM-modulation index and $m_\text{LI}$ is amplitude modulation index on the rf signal at $\tpd$. For the measurements shown in the main text $m_\text{LI}=1$. In this case one can write the field of the EOM output (cavity input) in the time domain as:

\begin{eqnarray}
 a_\textrm{out}(t)&=&a_o\Large[\cos(\omegap t) + \frac{m}{2}\left[\cos((\omegap+\tpd) t)+\cos((\omegap-\tpd)t)\right] \nonumber \\
   &+& \frac{m}{4}( \cos((\omegap+\tpd+\omega_\text{LI}) t) + \cos((\omegap+\tpd-\omega_\text{LI}) t)]\nonumber \\
   &+& \cos((\omegap-\tpd+\omega_\text{LI}) t) + \cos((\omegap-\tpd-\omega_\text{LI}) t) )].
\end{eqnarray}
Fig.~\ref{fig:exp_setup}(c) shows a schematic of the cavity input fields in the frequency domain. The reflected signal is
filtered by the cavity dispersion and considering the case where the pump is on the red-side of the cavity ($\omegap<\omega_o$) the reflected field is:

\begin{eqnarray}
    a_{R}(t)=r(\omega_s)\frac{a_o\alpha}{4}&\Large[\cos((\omegap+\tpd) t)+\cos((\omegap+\tpd)t +(\omega_\text{LI}t -\varphi)) + \cos((\omegap+\tpd)t -(\omega_\text{LI}t -\varphi))\Large]
\end{eqnarray}
First we assume that $r(\omega)$ is roughly constant over a range of $\omega_\text{LI}$ which is true for $\omega_\text{LI}<(\gamma_i+\gamma_\text{om})/2$. This implies that the smallest transparency window we could measure is on the order of the lock-in detection frequency, which corresponds to the smallest input power for the low and room temperature data.

We can now write the time average detected power spectral density on the photoreceiver (Det.~1 on Fig.~\ref{fig:exp_setup}a) by taking the absolute square value of the reflected field and keeping only the terms with frequency smaller than the detector bandwidth. In this case:
\bea \nonumber
    P|_{\omega_s} &=& \frac{a_o^2\alpha^2R_{PD}G_{PD}}{8R_L}|r(\omega_s)|^2\left[3+4\cos(\omega_\text{LI}t-\varphi)+\frac{1}{2}\cos(2\omega_{LI}t-2\varphi)+ \mathcal{O}(2\omegap)]\right].
\eea
where $R_{PD}=0.6$~A/V is the detector responsivity, $G_{PD}=1000$~V/A is the detector gain and $R_L=50~\Omega$ is the load resistance.

This signal is then sent to the lock-in which can measure independently the in-phase ($X$) and quadrature ($Y$) power spectral densities at $\omega_{LI}$:
\bea \nonumber
    X|_{\omega_\text{LI}} &=& \frac{a_o^2\alpha^2R_{PD}G_{PD}}{4R_L}|r(\omega_s)|^2\cos(\varphi)\\
    Y|_{\omega_\text{LI}} &=& \frac{a_o^2\alpha^2R_{PD}G_{PD}}{4R_L}|r(\omega_s)|^2\sin(\varphi)\\
\eea
It is then easy to see the reflection amplitude and phase are given by
\bea \nonumber
    |r(\omega_s)|^2=\frac{4R_L}{a_o^2\alpha^2R_{PD}G_{PD}}\sqrt{X|_{\omega_\text{LI}}^2+Y|_{\omega_\text{LI}}^2}\qquad\text{and}\qquad
    \tan(\varphi) = \frac{Y|_{\omega_\text{LI}}}{X|_{\omega_\text{LI}}}.
\eea
From the imparted change in the phase the signal delay is then calculated as:
\bea \nonumber
    \tau^{\text{(R)}} = \frac{\varphi}{\omega_\text{LI}}
\eea
where $\tau^{\text{(R)}}>0$ ($\tau^{\text{(R)}}<0$) represent a delay (advance) on the signal.

Here we have neglected the gain provided by the lock-in, which is important to determine the absolute value of $r(\omega_s)$. To account for that we calibrate the $X$ channel by a normalized transmission curve taken with low input power. Our assumption is that the cavity-taper coupling is not affected by the input power. A analogous result can be found for the case where the control laser is on the blue side of the cavity ($\omegap>\omega_o$).

\section{Room-temperature Electromagnetically Induced Transparency}

\begin{figure}[ht!]
\begin{center}
\includegraphics[width=7cm]{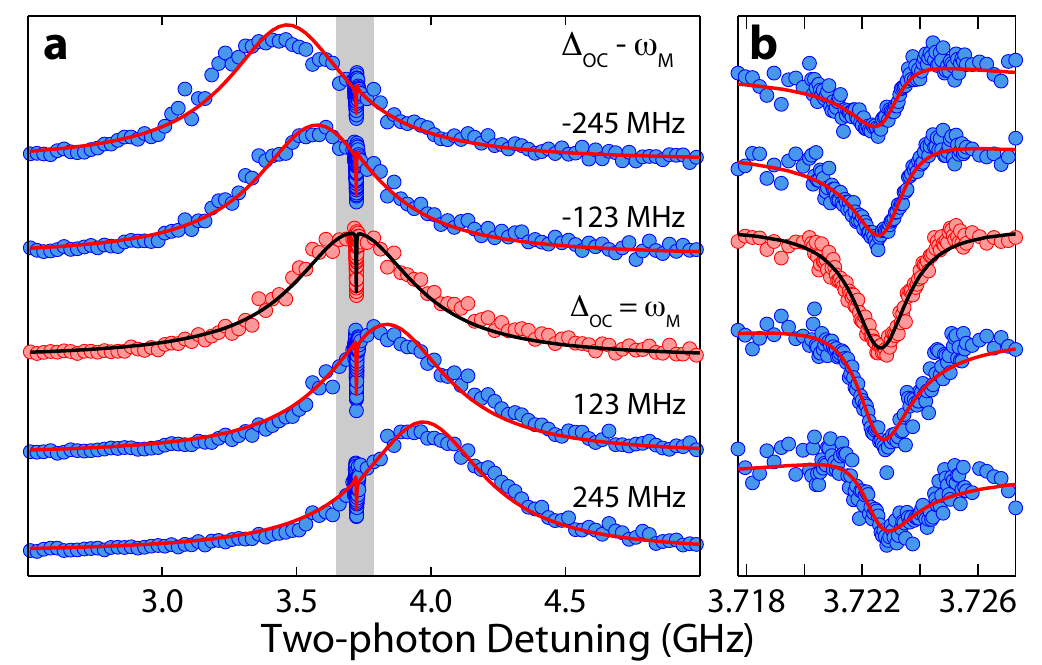}
\end{center}
\caption{\textbf{Room-temperature and amplification results. a,} room temperature normalized reflection signal from the probe laser as a function of the two photon detuning. Each curve represents a different control laser detuning from the optical cavity ($\DeltaOP$). The transparency window data as well as the fitted curve is shown around the mechanical frequency in \textbf{b}. \label{fig:RT}}
\end{figure}

Reflection spectroscopy of the system at 296 K results in the spectra shown in Fig.~\ref{fig:RT}. Due to the larger intrinsic mechanical damping rate at room temperature ($\gamma_i=2\pi\times1.9~\text{MHz}$), higher power is required to reach a given cooperativity. Additionally, the nanobeam optomechanical system is thermally sensitive and responds at a rate faster than the 89 kHz modulation signal used. As such, part of the phase response is thermal in nature. This added effect masks the small coherent phase-shifts imparted by the optomechanical cavity on the modulated signal sidebands.  Nonetheless, the Fano resonances measured (See Fig.~\ref{fig:RT}b) are a direct indication of coherent interference between the excitation of the optical cavity and the mechanical phonon.

\bibliographystyle{naturemag}

\end{document}